\documentclass[fleqn,usenatbib]{mnras}
\usepackage{newtxtext,newtxmath}
\usepackage[T1]{fontenc}
\usepackage{multirow}
\DeclareRobustCommand{\VAN}[3]{#2}
\let\VANthebibliography\thebibliography
\def\thebibliography{\DeclareRobustCommand{\VAN}[3]{##3}\VANthebibliography}
\usepackage{graphicx}
\usepackage{amsmath}
\usepackage{xcolor}
\usepackage{fix-cm}

\title[SSA Effects on Polarization Diagnostics]{Effects of Synchrotron Self-Absorption on Polarization Diagnostics of MHD Turbulence}

\author[N.-N. Guo et al.]{Ning-Ning Guo,$^{1}$
Jian-Fu Zhang,$^{1,2,3}$\thanks{E-mail: jfzhang@xtu.edu.cn (JFZ)} Hua-Ping Xiao,$^{1,3}$\thanks{E-mail: hpxiao@xtu.edu.cn (HPX)} and Qing-Zhong Liu$^{1,3}$\thanks{E-mail: qzliu@pmo.ac.cn (QZL)}
\\
$^{1}$Department of Physics, Xiangtan University, Xiangtan, Hunan 411105, China\\
$^{2}$National Astronomical Observatories, Chinese Academy of Sciences, 20A Datun Road, Chaoyang District, Beijing 100012, China\\
$^{3}$Key Laboratory of Stars and Interstellar Medium, Xiangtan University, Xiangtan 411105, China}

\date{Accepted XXX. Received YYY; in original form ZZZ}
 
\pubyear{\the\year{}}

\begin{document}
\label{firstpage}
\pagerange{\pageref{firstpage}--\pageref{lastpage}}
\maketitle

\begin{abstract}
Synchrotron polarization is a powerful tool for diagnosing magnetohydrodynamic (MHD) turbulence, yet prior analyses are confined to the optically thin regime, neglecting the effects of synchrotron self-absorption (SSA). Here, we incorporate SSA into the radiative transfer of polarized synchrotron emission and employ 3D MHD turbulence simulations to investigate its impacts on polarization statistics. We show that SSA strongly modulates polarized intensity by suppressing its emission amplitude, steepening its low-frequency power-law slopes, and reducing scale-dependent anisotropy. Notably, SSA does not impede the recovery of the scaling slope of the magnetic turbulence. Our results extend the applicability of synchrotron polarization diagnostics to optically thick astrophysical environments, providing a theoretical foundation for interpreting multi-frequency radio polarization observations. This confirms that polarization statistical techniques remain effective for diagnosing MHD turbulence properties in astrophysical environments where SSA is present.
\end{abstract}


\section{Introduction}
\label{Sec_Intro}
MHD turbulence is ubiquitous in the interstellar medium (ISM) and plays a crucial role in a wide range of astrophysical processes \citep[see][for a review]{Burkhart2021}. 
Synchrotron radiation serves as a powerful probe of Galactic magnetic fields and cosmic rays, and its polarization properties are extensively used to investigate the structure of interstellar magnetic fields, the characteristics of turbulence, and the propagation of cosmic rays (\citealt{Rybicki1979,Waelkens2009,Beck2015}). 

Theoretical investigations based on synchrotron radiation theory (\citealt{Rybicki1979}) have established a close relationship between synchrotron polarization statistics and the properties of MHD turbulence (\citealt{Waelkens2009,LP12,LP16}). Supported by subsequent numerical simulations and observational tests, statistical techniques rooted in synchrotron polarization have matured into robust tools for characterizing magnetized turbulence in ISM. For instance, \citet{Zhang2019a,Zhang2019b} inferred magnetic field orientations from the polarized intensity gradient; \citealt{Lazarian2018} and \citet{Guo2024} derived turbulent magnetic field strengths using the polarized intensity gradient and polarization variance, respectively; \citet{Zhang2025} reconstructed the spectral index of MHD turbulence via synchrotron polarization gradients; and \citet{Wang2020,Wang2024} explored turbulence anisotropy and intermittency through polarized intensity statistics. As such, synchrotron polarization has emerged as a key diagnostic for investigating magnetized turbulence in the ISM.

In a magnetized plasma, Faraday rotation (FR) induces a wavelength-dependent rotation of the polarization angle \citep{Burn1966, Brentjens2005}, which gives rise to a frequency-dependent polarized intensity and introduces systematic differences in the polarized intensity power spectra across observing bands (e.g., \citealt{Zhang2018}). Building on this frequency dependence, \citet{LP16} developed the Polarization Frequency Analysis (PFA) technique, which connects polarization statistics at distinct frequencies to both Faraday rotation and the spatial scales of turbulent fluctuations. By examining the frequency-dependent evolution of polarization statistics, PFA enables disentangling the contributions of different turbulent scales to the observed polarization signal, and thus extracting the scaling properties of MHD turbulence. Validated further by targeted numerical simulations from \citet{Zhang2016} and \citet{Xiao2025}, this method has been applied to three-dimensional Galactic datasets to investigate the scaling slopes of the Galactic disk and halo \citep{Xiao2026}.

We note that most existing statistical techniques for studying polarization have primarily focused on the effects of FR on magnetic turbulence, derived under the assumption of optically thin synchrotron emission. Consequently, the existing methods often overlook the role of SSA. In reality, SSA could significantly alter the frequency dependence of both total intensity and polarization parameters in the centimeter- to decimeter-wavelength regime (e.g., \citealt{Melrose1972,Jones1977}). Therefore, nearly all current statistical methods fail to account for the potential effects introduced by SSA. This oversight can lead to an underestimation or misinterpretation of the turbulent polarization properties.

The primary motivation of this work is to assess how SSA influences polarization statistics. Specifically, we aim to elucidate SSA's impact on the frequency dependence, spatial characterization, and turbulence diagnostics derived from optically thin polarization methods, thereby extending the applicability of these techniques for probing astrophysical magnetic fields. The paper is structured as follows: Section \ref{Sec_Theory} outlines the theoretical framework of synchrotron emission and SSA; Section \ref{Sec_Data} briefly describes the generation of three-dimensional MHD turbulence and synthetic polarization datasets; Section \ref{Sec_Result} presents our primary results; Section \ref{Sec_Diss} provides the discussion, following a summary in Section \ref{Sec_Sum}.

\section{Theoretical Foundations} 
\label{Sec_Theory}
Synchrotron emission originates from the helical motion of ultrarelativistic electrons embedded in a magnetic field, and SSA proceeds simultaneously \citep{Ginzburg1965,Rybicki1979}. These emission and absorption processes jointly control the evolution of the radiation field, which we describe via the polarized radiative transfer equation \citep{Jones1977}:
\begin{equation}
\begin{pmatrix}
\left(\frac{d}{dl}+a_I\right) & a_Q & a_U & a_V \\
a_Q & \left(\frac{d}{dl}+a_I\right) & r_V & -r_U \\
a_U & -r_V & \left(\frac{d}{dl}+a_I\right) & r_Q \\
a_V & r_U & -r_Q & \left(\frac{d}{dl}+a_I\right)
\end{pmatrix}
\begin{pmatrix}
I \\ Q \\ U \\ V
\end{pmatrix}
=
\begin{pmatrix}
j_I \\ j_Q \\
j_U \\ j_V
\end{pmatrix},
\label{Eq_transfer}
\end{equation}
where $I$ denotes the total specific intensity, $Q$ and $U$ quantify linear polarization, and $V$ describes circular polarization. The derivatives $d(I,Q,U,V)/dl$ track variations along the ray path. Eq.~(\ref{Eq_transfer}) incorporates synchrotron emission, absorption, and the evolution of polarization states during propagation. The coefficients $j_{I,Q,U,V}$ correspond to polarized emissivities, $a_I$ gives the total absorption coefficient, and $a_{Q,U,V}$ describe polarization-dependent absorption. The coefficient $r_V$ controls Faraday rotation and couples $Q$ and $U$, which rotates the linear polarization angle. The coefficients $r_Q$ and $r_U$ describe Faraday conversion, which couples linear and circular polarization.

We assume an isotropic pitch-angle distribution for relativistic electrons, whose energy distribution follows a power-law form $N(\gamma)d\gamma=\kappa\gamma^{-p}d\gamma$. Here, $N(\gamma)d\gamma$ denotes the number density of electrons with Lorentz factors between $\gamma$ and $\gamma+d\gamma$, $p$ is the electron power-law index, and $\kappa$ is the normalization constant. For a prescribed electron kinetic energy density $E_k$, $\kappa$ takes the following expressions. For $p\neq2$:
\begin{equation}
\kappa=\frac{E_k}{mc^2}
\left[
\frac{\gamma_{\rm max}^{2-p}-\gamma_{\rm min}^{2-p}}{2-p}
-\frac{\gamma_{\rm max}^{1-p}-\gamma_{\rm min}^{1-p}}{1-p}
\right]^{-1},
\end{equation}
and for $p=2$:
\begin{equation}
\kappa=\frac{E_k}{mc^2}
\left[\ln\left(\frac{\gamma_{\rm max}}{\gamma_{\rm min}}\right)
+\gamma_{\rm max}^{-1}-\gamma_{\rm min}^{-1}\right]^{-1}.
\end{equation}
Here, $m$ is the electron mass, $c$ is the speed of light, and $\gamma_{\rm min}$ and $\gamma_{\rm max}$ denote the lower and upper bounds of the electron Lorentz factor distribution, respectively. We set $\gamma_{\rm min}=10$ and $\gamma_{\rm max}=10^6$ in all subsequent calculations.

Based on the electron distribution described above, we obtain the synchrotron emission coefficients
\begin{equation}
j_I = j_{\alpha}j_{\perp}(\nu_{B\perp}/{\nu})^{\alpha}, \quad j_Q = j_{\alpha}^{Q}j_{\perp}(\nu_{B\perp}/{\nu})^{\alpha},
\end{equation}
together with the corresponding absorption coefficients
\begin{equation}
a_I = a_{\alpha}a_{\perp}(\nu_{B\perp}/{\nu})^{\alpha+5/2},
\quad a_Q = a_{\alpha}^{Q}a_{\perp}(\nu_{B\perp}/{\nu})^{\alpha+5/2},
\label{Eq_ab}
\end{equation}
where $j_{\perp} = (m c^2)(r_e/c)\nu_{B\perp}\kappa$ and $a_{\perp} = (r_e c)\nu_{B\perp}^{-1}\kappa$, with $r_e=e^2/(mc^2)$ being the classical electron radius. Here, $\alpha=(p-1)/2$ is the spectral index, $\nu$ is the emission frequency, and $\nu_{B\perp}=eB_{\perp}/(2\pi mc)$ is the perpendicular gyrofrequency. The quantities $e$ and $B_{\perp}$ denote the electron charge and the plane-of-sky magnetic field, respectively. The dimensionless functions $j_{\alpha}$, $j_{\alpha}^{Q}$, $a_{\alpha}$, and $a_{\alpha}^{Q}$ depend solely on the electron spectral index and are given by
\begin{align}
j_{\alpha}
= \frac{3^{\alpha+1/2}}{4(\alpha+1)}
\Gamma\left(\frac{\alpha}{2}+\frac{11}{6}\right)
\Gamma\left(\frac{\alpha}{2}+\frac{1}{6}\right),    
\end{align}
\begin{align}
j_{\alpha}^{Q}
= \frac{3^{\alpha+1/2}}{4(\alpha+5/3)}
\Gamma\left(\frac{\alpha}{2}+\frac{11}{6}\right)
\Gamma\left(\frac{\alpha}{2}+\frac{1}{6}\right),
\end{align}
\begin{align}
a_{\alpha}
= \frac{3^{\alpha+1}}{4}
\Gamma\left(\frac{\alpha}{2}+\frac{25}{12}\right)
\Gamma\left(\frac{\alpha}{2}+\frac{5}{12}\right),
\end{align}
\begin{align}
a_{\alpha}^{Q}
= \frac{3^{\alpha+1}}{4}
\frac{\alpha+3/2}{\alpha+13/6}
\Gamma\left(\frac{\alpha}{2}+\frac{25}{12}\right)
\Gamma\left(\frac{\alpha}{2}+\frac{5}{12}\right).
\end{align}
All radiative transfer coefficients are expressed in CGS units. The SSA optical depth is given by $\tau=\int a_I(l,\nu)\,dl$. The limits $\tau \ll1$ and $\tau \gg1$ correspond to the optically thin and optically thick regimes, respectively. Each computational cell is assumed locally homogeneous. In each cell, the polarization reference axes are chosen parallel and perpendicular to the projected magnetic field. Positive $Q$ corresponds to an electric vector perpendicular to the projected field. Within this local frame, $j_U=0$, $a_U=0$, and $r_U=0$. Since this work focuses on linear polarization, we neglect intrinsic circularly polarized emission ($j_V=0$), circular dichroism ($a_V=0$), and Faraday conversion ($r_Q=0$), while retaining Faraday rotation.

The Faraday rotation coefficient receives contributions from both relativistic electrons and nonrelativistic  background thermal electrons \citep{Jones1977}:
\begin{equation}
r_V = r_{\alpha}^{V}a_{\perp} (\nu_{B\perp}/{\nu})^2
(\ln\gamma_{\rm min})\gamma_{\rm min}^{-2\alpha-2}\cot\theta_B+r_V^{(\mathrm c)},
\end{equation}
where $\theta_B$ is the angle between the local magnetic field and the propagation direction, such that $B_{\perp}=B\sin\theta_B$ and $B_{\parallel}=B\cos\theta_B$. The dimensionless coefficient reads $r_{\alpha}^{V}=(2\alpha+3)/(\alpha+1)$. The background electron contribution is given by:
\begin{equation}
r_V^{(\mathrm c)}=\frac{2r_ec n_{\mathrm c}}{\nu_{B\perp}}
(\nu_{B\perp}/{\nu})^2\cot\theta_B,
\end{equation}
where $n_{\mathrm c}$ denotes the number density of background electrons. The corresponding rotation measure (RM) is defined as
\begin{equation}
{\rm RM}(\boldsymbol{X}, z) = 0.812 \int_{0}^{z} \left(\frac{n_c}{\rm{cm}^{-3}}\right) \left(\frac{B_\parallel}{\mu\rm{G}}\right) \frac{dl}{\rm{pc}} \ \rm{rad\ m}^{-2}.
\end{equation}
Here, $\boldsymbol X$ denotes the position in the plane of the sky, the $z$-axis is aligned with the line of sight (LOS), with $B_{\parallel}$ positive when pointing toward the observer. The path coordinate $l$ increases toward the observer, from $l=0$ to $l=L$ across the emitting region.

With these coefficients, we solve Eq. (\ref{Eq_transfer}) along the LOS to obtain the Stokes parameters $I$, $Q$, and $U$. We then compute the linearly polarized intensity $P$, the linear polarization fraction $\Pi_L$, and the polarization angle $\psi$ from
\begin{equation}
P=\sqrt{Q^2+U^2},\quad \Pi_L=\frac{P}{I},\quad \psi=\frac{1}{2}\tan^{-1}\left(\frac{U}{Q}\right).
\end{equation}
The quadrant of the polarization angle is determined by the signs of $Q$ and $U$, and we compute $\psi$ numerically via the two-argument arctangent function. This angle encapsulates the joint effects of intrinsic polarized synchrotron emission, SSA, and Faraday rotation along the LOS. To disentangle the respective roles of SSA and Faraday rotation, we define two complementary test cases with identical magnetic configurations, electron distributions, and emission coefficients. The FR-only case turns off absorption by setting $a_I = 0$ and $a_Q = 0$ while preserving $r_V$, whereas the SSA-only case retains all absorption terms and sets $r_V = 0$.

\section{Data Description} 
\label{Sec_Data}
To synthesize synchrotron polarization observations, we employed four sets of three-dimensional quasi-steady data cubes characterizing the magnetized ISM, where MHD equations together with an isothermal equation of state are solved using the second-order-accurate hybrid nonoscillatory code \citep{Cho2003}. Upon obtaining these datasets, we adopted the following configurations: a periodic box\footnote{Periodic boundary conditions apply only to the MHD simulations, while the radiative transfer calculation traverses the box once along each LOS, without periodic wrapping.} with a fixed size of $2\pi$, a variable mean magnetic field $B_0$ oriented along the $x$-axis, a variable gas pressure $p_0$, an external driving mode at a fixed wave number of $k \approx 2.5$, and a numerical resolution of $512^3$ grid points. When a simulation reaches a quasi-steady state—determined by the stable distribution of the power spectrum—we characterize the properties of the 3D datasets using two key parameters: the Alfv\'{e}n Mach number $M_{\rm A}$ and the sonic Mach number $M_{\rm s}$.

To generate synchrotron polarization observables, we first convert the dimensionless density and magnetic field from the three-dimensional MHD data cubes into dimensional physical quantities with realistic units. We adopt a thermal electron density $n_c=0.1\ \rm cm^{-3}$ and magnetic field strength $B=10\ \rm mG$. We further set the non-thermal electron spectral index to $p=2.7$ and compute the synchrotron emission, absorption, and Faraday rotation coefficients within each grid cell following the formalism described in Section \ref{Sec_Theory}. Finally, we solve the polarized radiative transfer equation cell by cell along the line of sight to obtain the Stokes parameters $I$, $Q$, and $U$, from which the polarized intensity $P$ is derived. All calculations use a line-of-sight integration depth $L=100\ \rm pc$, matching the full extent of the simulation domain along the line of sight, except for Fig. \ref{fig:parameters}, which explicitly explores parameter dependencies of SSA. For the dataset with $M_{\rm A}=0.65$ and $M_s=0.48$, the RM distribution integrated over the full line of sight yields a mean of $86.1\ \rm rad\ m^{-2}$ and a standard deviation of $1515.7\ \rm rad\ m^{-2}$. As the frequency rises from $0.1$ to $10\ \rm GHz$, the mean total optical depth decreases from $683.2$ to $1.4\times 10^{-4}$.

\section{Results} 
\label{Sec_Result}
\subsection{Parameter Dependence of SSA and Intensity Spectral Energy Distributions}
\label{Subsec_nu}
\begin{figure*}
    \includegraphics[width=\textwidth]{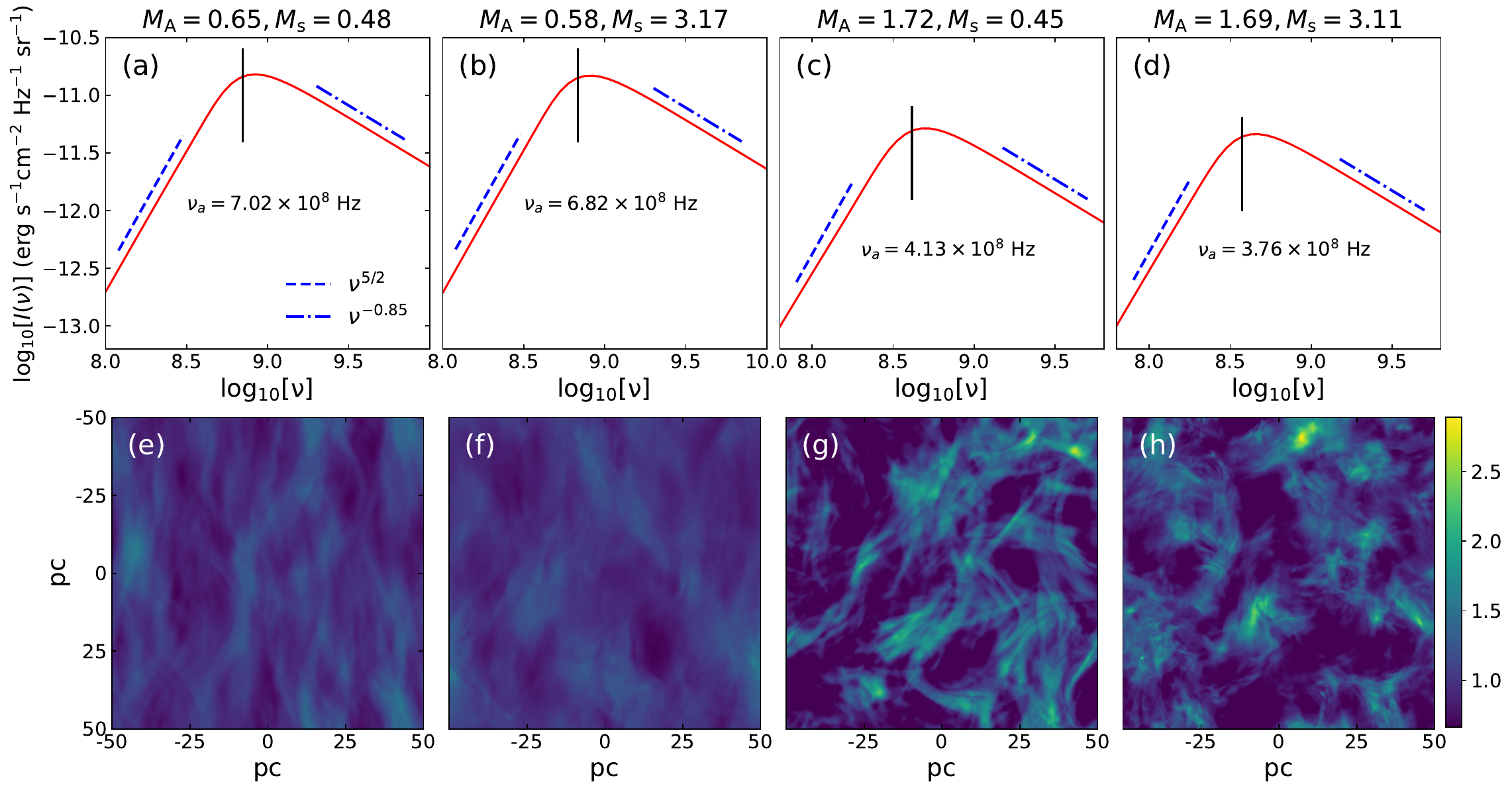}
    \caption{Basic characteristics of synchrotron radiation self-absorption in different MHD turbulent regimes. Top row: Spectral energy distributions of synchrotron intensity. Bottom row: Spatial distribution of self-absorption optical depth at the SSA turnover frequency}.
    \label{fig:ab_nu}
\end{figure*}

In this section, we investigate how key relevant parameters influence the SSA effect under the assumption of negligible Faraday depolarization, and how such an effect modifies the spectral energy distribution of the synchrotron intensity. To understand how SSA affects polarization statistics, it is first necessary to determine the SSA turnover frequency range and analyze the spatial distribution of absorption coefficients. The upper panels of Fig. \ref{fig:ab_nu} show the variation of the total synchrotron intensity $I(\nu)$ as a function of frequency $\nu$. Consistent with theoretical expectations, in the low-frequency regime where $\tau \gg 1$, the intensity follows $I(\nu) \propto \nu^{5/2}$, whereas in the high-frequency regime where $\tau \ll 1$, $I(\nu) \propto \nu^{-(p-1)/2} = \nu^{-0.85}$, where setting $p=2.7$. The SSA turnover frequency $\nu_{\rm a}$, labeled in the figure, corresponds to the frequency at which $I(\nu)$ peaks.

The lower panels of Fig. \ref{fig:ab_nu} show the distributions of the optical depth at the SSA turnover frequency for the four data sets. In sub-Alfv\'enic (strongly magnetized, $M_{\rm A}<1$) turbulence, the optical depth exhibits a more uniform spatial distribution with suppressed fluctuations. In contrast, in super-Alfv\'enic (weakly magnetized, $M_{\rm A}>1$) turbulence, the optical depth shows prominent filamentary and cavity-like structures with significantly larger variations. This is because the optical depth distribution is determined by the absorption coefficient, whose spatial variation is governed by the magnetic field structure for a fixed spectral index and frequency (see Eq. (\ref{Eq_ab})). Since $M_{\rm A}\propto \delta B/B_0$ (\citealt{Lazarian2025}, where $\delta B$ denotes the turbulent magnetic field fluctuation), the sub-Alfv\'enic turbulence regimes are dominated by the ordered mean field and therefore exhibit weaker optical depth fluctuations, whereas the super-Alfv\'enic turbulence regimes are dominated by turbulent magnetic fluctuations and show stronger spatial variations.

\begin{figure*}
    \includegraphics[width=\textwidth]{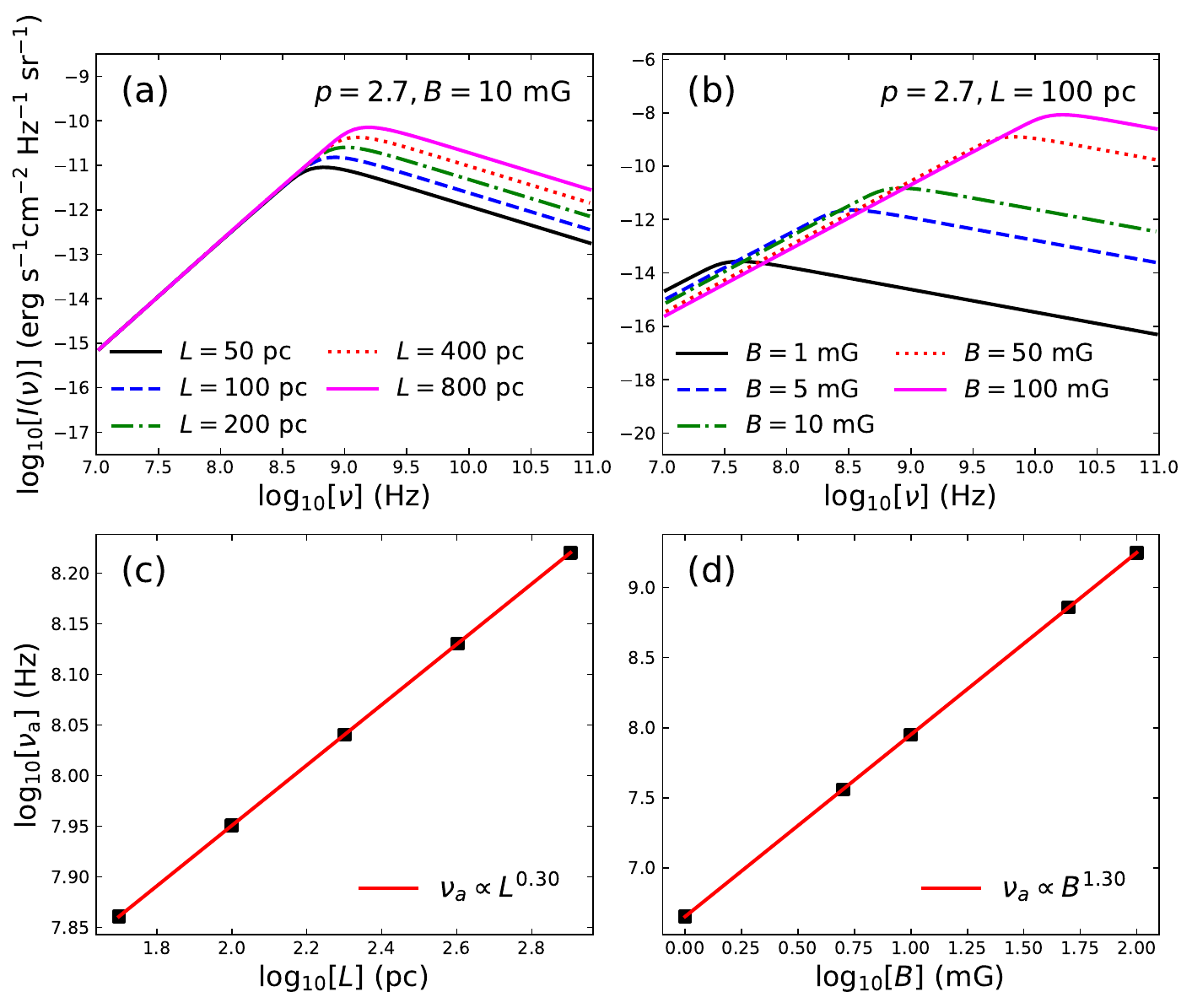}
    \caption{Top row: Effects of integration depth $L$ and magnetic field strength $B$ on the spectral energy distributions. Bottom row: The SSA turnover frequency $\nu_a$ plotted as functions of $L$ and $B$.}
    \label{fig:parameters}
\end{figure*}

In Fig. \ref{fig:parameters}, we explore the effects of integration depth and magnetic field strength on the SSA turnover frequency $\nu_a$. The upper panels show that as $L$ and $B$ increase, the peak values of $I(\nu)$ increase and $\nu_a$ shifts to higher frequencies, indicating that $\nu_a$ is positively correlated with both parameters. The lower panels plot $\nu_a$ as a function of $L$ and $B$. We find that $\nu_a$ follows power-law relations of the form $\nu_a \propto L^{0.30}$ and $\nu_a \propto B^{1.30}$ over the explored parameter range. These scalings are consistent with the theoretical prediction $\nu_a \propto (\kappa B_{\perp}^{(p+2)/2} L)^{2/(p+4)}$, where the normalized constant $\kappa$ is associated with magnetic energy $B^2/8\pi$; the expected scalings are $\nu_a \propto L^{0.3}$ and $\nu_a \propto B^{1.3}$ for $p = 2.7$.

\subsection{Depolarization Effects on Polarization Spectral Distributions and Spatial Distribution}
In the process of radiation transfer, FR introduces a frequency-dependent depolarization effect that causes the polarization angle to rotate, thereby altering the spatial distribution of the polarized intensity. In contrast, SSA reduces the effective emission depth and modifies the formation region of the polarized signal, which leads to systematic changes in both the amplitude and spatial distribution of the polarized intensity.

\begin{figure*}
    \includegraphics[width=\textwidth]{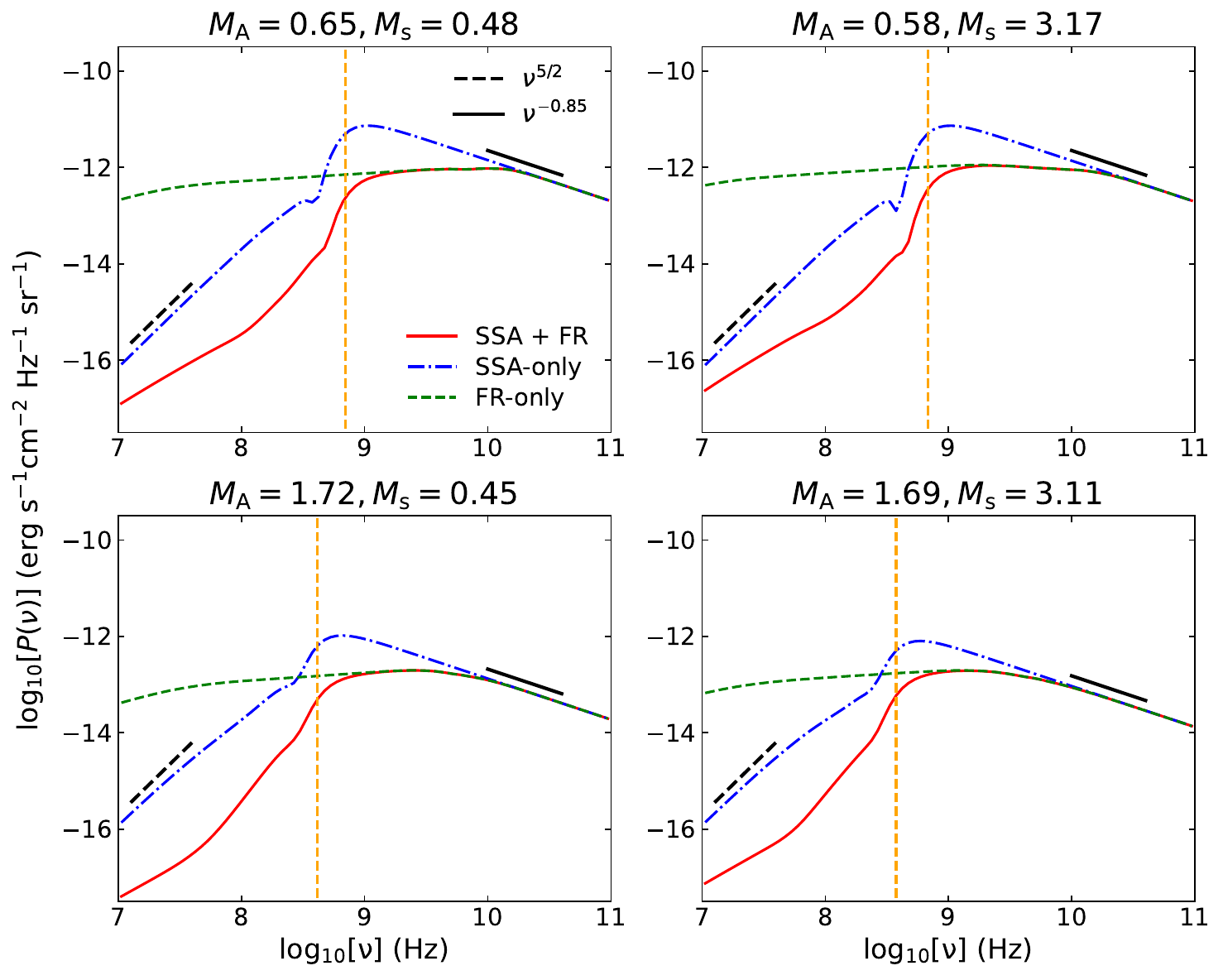}
    \caption{The spectral energy distribution of polarized intensity under different MHD turbulence regimes for three cases: SSA only, FR only, and the combined effects of SSA and FR. The vertical orange dashed line marks the SSA turnover frequency.}
    \label{fig:PI_nu}
\end{figure*}

Fig. \ref{fig:PI_nu} presents the spectral energy distributions of the synchrotron polarized intensity $P(\nu)$ versus frequency $\nu$ for four turbulence models with distinct Alfvén Mach numbers $M_{\rm A}$ and sonic Mach numbers $M_{\rm s}$.
At high frequencies the medium is optically thin, so SSA and Faraday depolarization remain negligible and all three curves converge to the intrinsic synchrotron scaling $P(\nu)\propto\nu^{-(p-1)/2}=\nu^{-0.85}$. The three cases diverge into distinct spectral shapes toward lower frequencies. The FR-only spectrum gradually flattens with decreasing frequency, because the Faraday rotation angle $\Delta\psi=\mathrm{RM}\lambda^2$ ($\lambda=c/\nu$) increases at longer wavelengths and enhances cancellation between polarized contributions along the line of sight, which suppresses the low-frequency rise of the intrinsic emission. For the SSA-only case, stronger self-absorption pulls the spectrum away from the optically thin $\nu^{-0.85}$ scaling at low frequencies. The polarized intensity rises to a peak and then falls off, approaching the asymptotic form $P(\nu)\propto\nu^{5/2}$ in the optically thick limit ($\tau\gg1$). The small dip near the SSA turnover frequency originates from a minimum of the polarization fraction accompanied by a reversal of polarization orientation.\footnote{\citet{Ginzburg1969} predicted that for a homogeneous source the two orthogonal linear polarization components satisfy $I_{\perp}>I_{\parallel}$ in the optically thin regime and $I_{\perp}<I_{\parallel}$ in the optically thick regime. At the frequency where $I_{\perp}=I_{\parallel}$, the linear polarization fraction $\Pi=|(I_{\perp}-I_{\parallel})/(I_{\perp}+I_{\parallel})|$ vanishes. Across this frequency the dominant polarization orientation reverses as the two components exchange dominance, creating a dip in the polarized-intensity spectrum, as supported numerically by \citet{Cowie2026}.} The SSA+FR spectrum closely tracks the FR-only curve where self-absorption is still weak. Further reduction of frequency causes the curves to separate and the low-frequency spectrum to steepen. Faraday depolarization governs the higher-frequency portion of this transition region, whereas the joint effect of SSA and Faraday depolarization sets the behavior at the lowest frequencies.

Comparing the four MHD turbulent regimes, a clear dependence on the magnetic field strength (quantified by $M_{\rm A}$) is observed, with sub-Alfv\'enic models exhibiting significantly higher peak polarized intensity and higher turnover frequencies $\nu_{\rm a}$ than super-Alfv\'enic models. The sonic Mach number $M_{\rm s}$ has a much weaker effect on the spectral shape, peak intensity, and turnover frequency for a given $M_{\rm A}$. These results confirm that SSA strongly modulates polarized synchrotron emission, with the magnetic field strength being the dominant parameter governing the spectral properties.

\begin{figure*}
    \includegraphics[width=\textwidth]{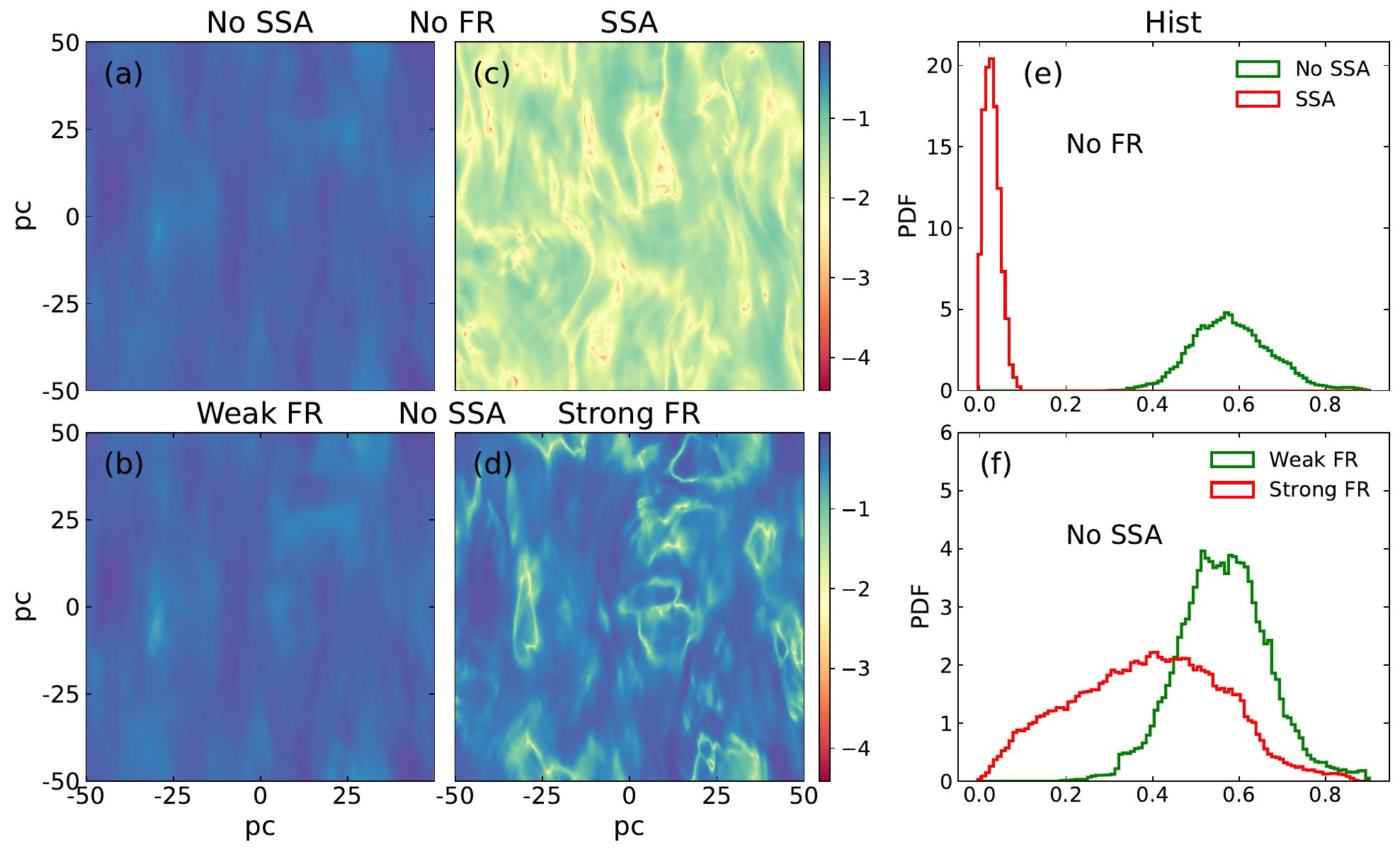}
    \caption{Impacts of SSA and FR on synchrotron polarization properties. Top row: Spatial maps of polarized intensity at $0.4\ \rm{GHz}$ without SSA (a) and with SSA (c) in the absence of FR, with the corresponding PDFs shown in panel (e). Bottom row: Spatial maps of polarized intensity without SSA under weak FR at $20\ \rm{GHz}$ (b) and strong FR at $10\ \rm{GHz}$ (d), with Faraday rotation angles ranging from $-0.95$ to $0.92\ \rm{rad}$ and from $-3.80$ to $3.70\ \rm{rad}$, respectively. The corresponding PDFs are shown in panel (f).}
    \label{fig:Map_PI}
\end{figure*}

Fig. \ref{fig:Map_PI} presents the spatial distributions of the synchrotron polarized intensity in normalized units and their probability distribution functions (PDFs), with separate comparisons of SSA and FR. Panels (a) and (c) show the polarized intensity distributions without and with SSA, respectively, at $0.4\ \rm{GHz}$ in the absence of FR. A clear suppression of polarized intensity is observed when SSA is included, with panel (c) showing a significantly lower overall intensity than the case without SSA. This suppression is further quantified in the PDF panel (e). The case without SSA shows a broad, peaked PDF centred at a high intensity value ($\sim 0.55$), whereas the case with SSA is strongly shifted towards much lower values, with a narrow, high-peaked distribution concentrated near $\approx 0.05$. Panels (b) and (d) compare the polarized intensity distributions under weak and strong FR without SSA, with their corresponding PDFs shown in panel (f). Strong FR significantly alters the spatial structure of polarized intensity, producing more complex filamentary and small-scale features. It also broadens the PDF and shifts its peak from $\sim 0.55$ to $\sim 0.4$.

These results demonstrate that SSA acts as a strong depolarization mechanism, drastically reducing the overall polarized emission level and modifying the statistical distribution of polarization intensity. Whereas FR mainly alters the spatial distribution of polarized intensity and locally suppresses its amplitude.

\subsection{Effects of SSA on Polarized Intensity Statistics}
\label{Subsec_Statist}
\subsubsection{Effects on Anisotropy}
\label{Subsubsec_Anisotropy}
\begin{figure*}
    \includegraphics[width=\textwidth]{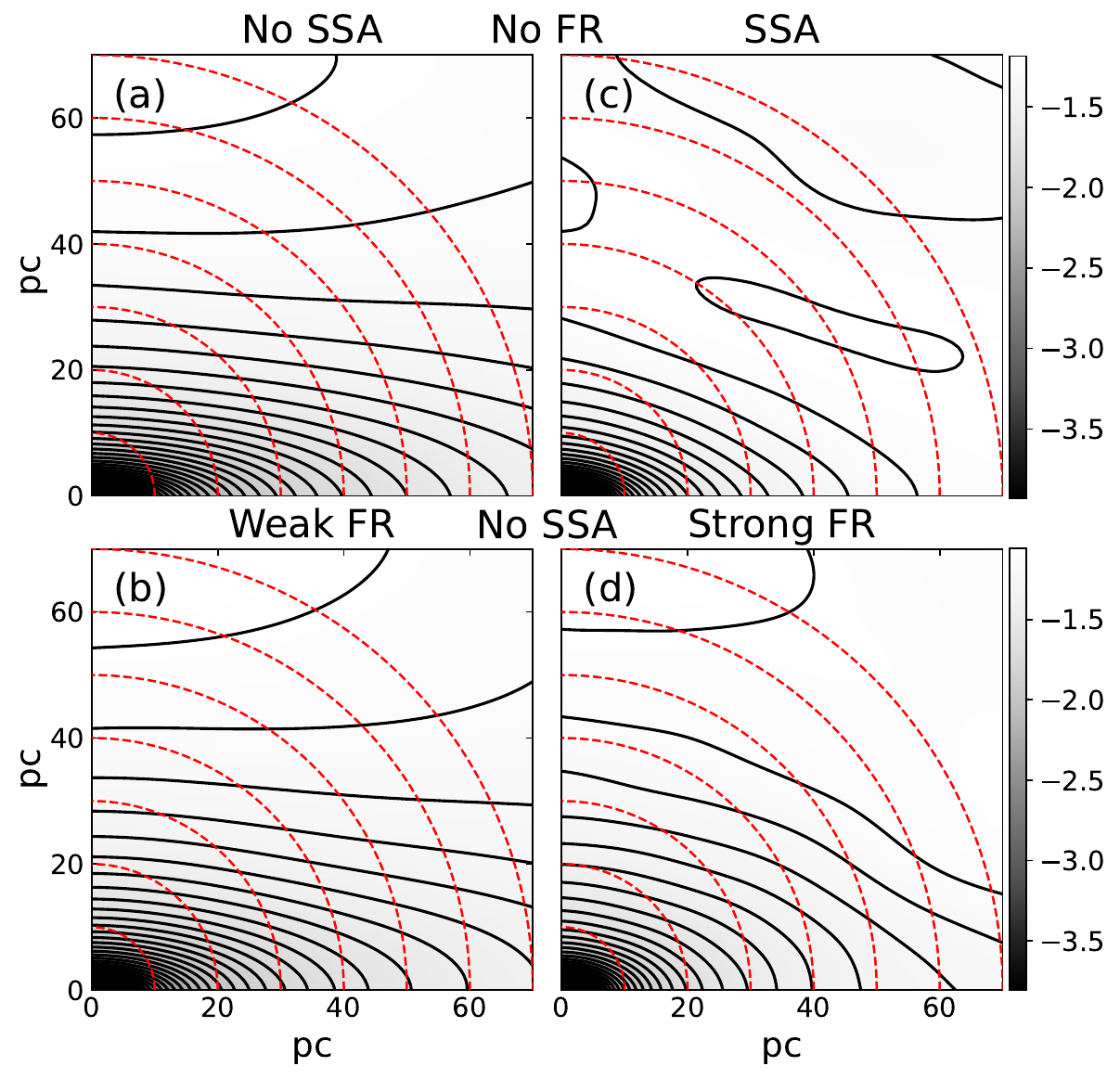}
    \caption{Impacts of SSA and FR on polarized intensity anisotropy under the same conditions as panels (a)--(d) of Fig. \ref{fig:Map_PI}. Black solid contours denote the intensity distribution anisotropy, with red dashed isotropic reference contours for comparison.} 
    \label{fig:SF}
\end{figure*}

Fig. \ref{fig:SF} presents contour maps of the synchrotron polarized intensity, illustrating the anisotropy of polarized emission under different conditions of SSA and FR. Panels (a) and (c) show the results without and with SSA, respectively, in the absence of FR. Panels (b) and (d) show the results under weak and strong FR without SSA, respectively. A clear anisotropic feature is observed in all cases, with the iso-intensity contours deviating from the circular reference lines. When SSA is included, the contours become less elongated and show a reduced deviation from the circular reference lines, indicating a clear reduction in the anisotropy of the polarized intensity distribution. A comparison between panels (b) and (d) shows that stronger FR also alters the contour morphology and slightly weakens the anisotropy of the polarized intensity.

These results suggest that SSA not only suppresses the overall level of polarized intensity but also strongly modulates its spatial anisotropy. FR also affects the anisotropic structure of polarized intensity, although its effect is weaker than that of SSA.

\subsubsection{Effects on the Power Spectrum}
\label{Subsubsec_Power}
\begin{figure*}
    \includegraphics[width=\textwidth]
    {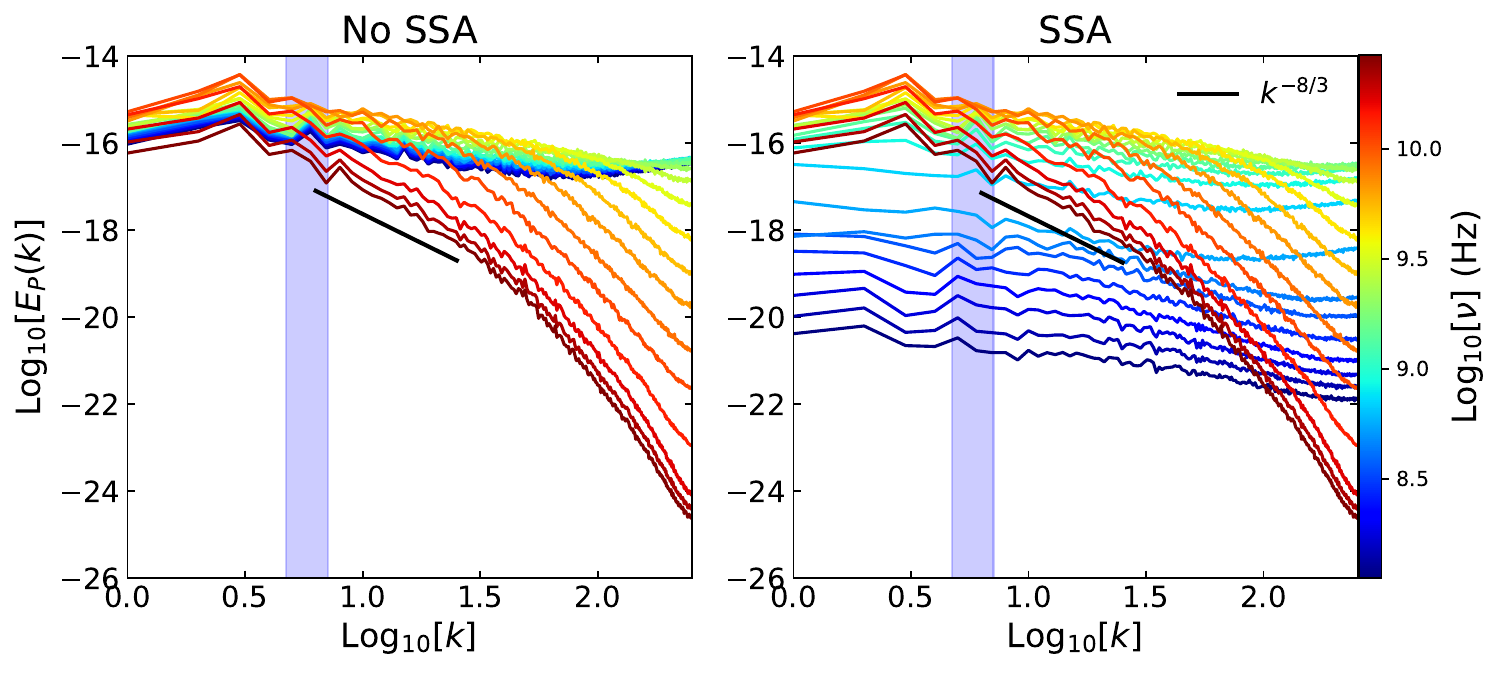}
    \caption{Polarization energy spectra $E_P(k)$ for the No SSA (left) and SSA (right) cases. The light purple regions denote the transition regions from weak to strong turbulence. Colored lines represent different frequencies.}
    \label{fig:Spectra}
\end{figure*}

Fig. \ref{fig:Spectra} presents the power spectra of synchrotron polarized intensity, comparing the cases with and without SSA. Each curve corresponds to a different observing frequency, color-coded by $\log_{10}[\nu]$ (in Hz) as shown in the right-hand color bar, spanning from $\sim 10^8$ Hz (dark blue, low frequency) to $\sim 3\times 10^{10}$ Hz (red, high frequency). The light purple shaded region marks the characteristic wavenumber range of the transition scale for the weak-to-strong turbulence.

The left panel shows the power spectra in the absence of SSA. At high frequencies, the spectra in the inertial range ($\log_{10}k \gtrsim 0.7$) follow the classic Kolmogorov $k^{-8/3}$ power-law scaling, consistent with the expected behavior of fully developed turbulence. At low wavenumbers (large scales, $\log_{10}k \lesssim 0.7$), the power spectra rise to a peak at the driving scale. As the frequency decreases, Faraday depolarization increasingly modifies the power spectra, causing them to deviate from the Kolmogorov scaling, particularly at large wavenumbers. The right panel displays the corresponding power spectra with SSA included. A dramatic modification of the spectral shape is observed: (1) At high frequencies, the Kolmogorov $k^{-8/3}$ scaling in the inertial range is preserved; (2) At low frequencies, the overall amplitudes of the power spectra are suppressed, the peaks near the driving scale become less pronounced, and the spectra become significantly flatter at larger scales.

As a result, the SSA acts as a strong filter on the polarized intensity power spectrum, drastically reducing the overall power while preserving the underlying turbulence scaling in the inertial range. The suppression effect is frequency-dependent, with stronger attenuation at lower frequencies, highlighting the critical role of SSA in shaping the observable polarization properties of turbulent astrophysical plasmas.

\subsubsection{Effects on the Polarization Variance and Wavelength Squared Relationship}
\label{Subsubsec_Variance}

\begin{figure*}
    \includegraphics[width=\textwidth]{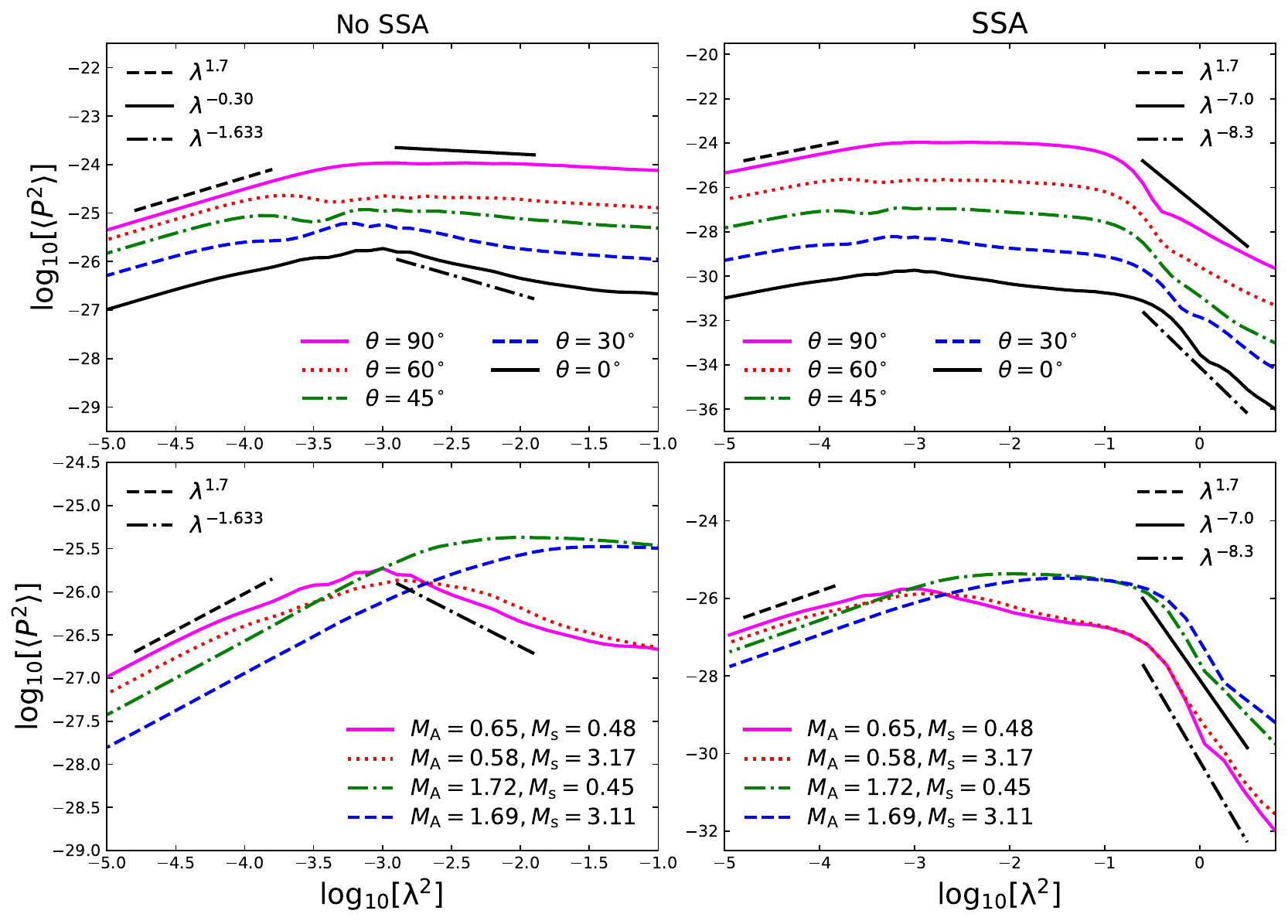}
    \caption{Upper panels: The variance of polarized intensity as a function of $\lambda^2$ for different angles between the LOS and the mean magnetic field direction (ranging from $\theta=90^\circ$ to $\theta=0^\circ$). Lower panels: The variance of polarized intensity as a function of $\lambda^2$ for four different turbulence regimes at $\theta=0^\circ$. Both the upper and lower panels are presented to compare the cases with and without SSA.}
    \label{fig:Var}
\end{figure*}

Fig. \ref{fig:Var} presents the power-law scaling of the polarized intensity variance $\langle P^2 \rangle$ as a function of wavelength squared $\lambda^2$, under different physical conditions with and without SSA. The upper row compares results for different LOS angles $\theta$ relative to the mean magnetic field, while the lower row shows results for four distinct turbulence models characterized by different Alfv\'en Mach numbers $M_{\rm A}$ and sonic Mach numbers $M_{\rm s}$. The left column displays the cases without SSA, and the right column shows the corresponding results with SSA included. The black dashed and dotted lines mark the theoretical power-law scalings.\footnote{In the optically thin regime, \citet{LP16} predicted that the power-law behavior of the polarized intensity variance as a function of wavelength squared $\lambda^2$ depends on the wavelength regime, the mean FR measure density $\overline{\Phi}$, and the root-mean-square (rms) FR measure density $\sigma_{\Phi}$. In the long-wavelength regime, when $\overline{\Phi} > \sigma_{\Phi}$, the variance scales as $\lambda^{-2-2m}$, where $m$ is the spectral index of magnetic turbulence; in contrast, when $\overline{\Phi} < \sigma_{\Phi}$, it follows a $\lambda^{-2}$ scaling. For the $\overline{\Phi} > \sigma_{\Phi}$ case, information about magnetic turbulence can be recovered from the derivative of $\lambda^2 \langle P^2 \rangle$ with respect to $\lambda^2$. In the short-wavelength regime, the variance scales as $\lambda^{-2+2m}$ when $\overline{\Phi} > \sigma_{\Phi}$, and as $\lambda^{(-2+2m)/(1-\tilde{m}_{\phi}/2)}$ when $\overline{\Phi} < \sigma_{\Phi}$, where $\tilde{m}_{\phi} = \min(m_{\phi}, 1)$ and $m_{\phi}$ is the correlation index of the FR measure density. These theoretical scalings have later been supported by numerical simulations \citep{Zhang2016,Xiao2025} and applied to observational data \citep{Xiao2026}. We stress that all these theoretical predictions, numerical validations, and observational applications are strictly for the optically thin regime.} We derive the power-law relations of the polarized intensity variance as a function of wavelength under different physical conditions, both with and without SSA, as summarized in Table \ref{Tab_Var}. 

In the absence of SSA (left panels), the variance follows the expected theoretical scalings: a rising power law $\propto \lambda^{1.7}$ at short wavelengths (weak FR) and a decaying power law $\propto \lambda^{-1.633}$ ($\bar{\phi} > \sigma_{\phi}$) or $\propto \lambda^{-0.30}$ ($\bar{\phi} < \sigma_{\phi}$) at long wavelengths (strong FR). The overall variance level increases with the LOS angle $\theta$, with the highest intensity observed at $\theta = 90^\circ$ and the lowest at $\theta = 0^\circ$. For different turbulence models, sub-Alfv\'enic models exhibit significantly higher variance than super-Alfv\'enic models. When SSA is included (right panels), the scaling of the polarization variance is significantly modified: At short wavelengths, the polarization variance follows the expected scaling $\propto \lambda^{1.7}$. At long wavelengths, SSA strongly suppresses the polarization variance and produces a considerably steeper decay, consistent with the theoretical scalings $\lambda^{-7.0}$ and $\lambda^{-8.3}$ ($m=2/3$). 

Quantitative analysis of the lower panels in Fig.\ref{fig:Var} reveals that the condition $\bar{\phi}>\sigma_{\phi}$ is satisfied only for the sub-Alfv\'enic, subsonic turbulence model with the LOS aligned with the mean magnetic field ($\theta=0^\circ$). Under this specific condition, the power-law scaling of the polarization variance exhibits an explicit dependence on the turbulence spectral index $m$, enabling the retrieval of the magnetic turbulence spectral index from the formulae listed in Table \ref{Tab_Var}. In contrast, for the remaining three turbulence models, the condition $\bar{\phi}<\sigma_{\phi}$ holds for all LOS angles relative to the mean magnetic field spanning $0^\circ$ to $90^\circ$. For these cases, the corresponding power-law relations of the polarization variance no longer include the index $m$, meaning the scaling behavior of the polarized intensity variance alone cannot be used to recover the magnetic turbulence spectral index. That said, following the approach of \cite{Xiao2025}—by considering the derivative of the polarized intensity variance with respect to wavelength squared (a topic beyond the scope of this work)—the spectral characteristics of magnetic turbulence can still be extracted even in such scenarios.

Despite the substantial modifications to polarization variance scaling induced by SSA, our results demonstrate that this relation still retains the key spectral signatures of the underlying magnetic turbulence, thus permitting the robust recovery of the turbulent power-law index. These findings confirm that while SSA reshapes the polarization variance spectra and modifies their scaling behavior, it does not preclude the extraction of fundamental magnetic turbulence properties from polarization observations.

\begin{table}
\centering
\caption{The power-law relationship between polarization variance ($P^2$) and wavelength under different conditions with and without SSA.}
\label{Tab_Var}
\begin{tabular}{l c c r}
\hline
 &  & No SSA & SSA \\
\hline
\multicolumn{2}{c}{Short $\lambda$} 
& $P^2\propto \lambda^{p-1}$ 
& $P^2\propto \lambda^{p-1}$ \\
\hline
\multirow{2}{*}{Long $\lambda$} 
& $\bar{\phi} > \sigma_{\phi}$ 
& $P^2\propto \lambda^{-2-2m+(p-1)}$ 
& $P^2\propto \lambda^{-2-2m-5}$ \\
& $\bar{\phi} < \sigma_{\phi}$ 
& $P^2\propto \lambda^{-2+(p-1)}$ 
& $P^2\propto \lambda^{-2-5}$ \\
\hline
\end{tabular}
\end{table}

\section{Discussion} 
\label{Sec_Diss}
In this paper, we explore the influence of SSA effects on recently developed polarized intensity statistical techniques. As an inherent depolarization mechanism in optically thick media, self-absorption has the most significant effect on reducing the polarization degree. For instance, considering an electron spectral index of $p=2.7$, the polarization degree is reduced from $\Pi\simeq 0.74$ in the optically thin case to 0.1 in the optically thick case. Similar to Faraday depolarization (\citealt{Burn1966,Brentjens2005}), our numerical results show that the inclusion of SSA significantly modifies the response of polarization statistics to the underlying turbulence properties—specifically, the polarized intensity is substantially reduced in amplitude and exhibits additional frequency dependence. Interestingly, we find that polarization statistics incorporating the SSA effect are not suppressed in revealing the spectral properties of the turbulence. Therefore, our study extends the application of correlated polarization statistical methods to astrophysical environments affected by SSA.

In general, the SSA is commonly observed in compact radio sources—characterized by high electron number densities and strong magnetic fields—and is particularly prominent at low radio frequencies. This physical effect significantly alters both the spectral energy distribution and polarization properties of synchrotron radiation, two key observational signatures of relativistic particle populations in astrophysical plasmas, and thus cannot be neglected in relevant astrophysical investigations. The parameter range adopted in this work broadly encompasses typical values inferred from astronomical observations of various compact radio source environments, including the hotspot regions of the powerful radio galaxy Cygnus A \citep{Carilli1991}, the interaction regions between radio supernova remnants and their surrounding circumstellar media—where shock waves accelerate relativistic electrons and generate strong magnetic fields, leading to prominent SSA signatures \citep{Chevalier1998}, X-ray binaries with relativistic jets that emit strong synchrotron radiation and exhibit clear SSA effects at low frequencies \citep{Corbel2002}, and the relativistic jets of active galactic nuclei \citep{Sullivan2009}. This comprehensive coverage of observational constraints thereby strongly supports the physical applicability and generality of our numerical results.

To simplify the calculations, we assume an isotropic pitch-angle distribution for relativistic electrons and a power-law distribution for their energy.\footnote{For a general electron energy distribution, the degree of linear polarization is obtained by computing the total emissivities in two orthogonal polarization directions, $J_\perp(\nu)$ and $J_\parallel(\nu)$, such that $\Pi(\nu)=\frac{J_\perp(\nu)-J_\parallel(\nu)}{J_\perp(\nu)+J_\parallel(\nu)}$ \citep{Rybicki1979}.
} Although in realistic astrophysical environments, the electron momentum distribution may exhibit anisotropy and the energy distribution may deviate from a pure power law, these idealized assumptions do not compromise the robustness of our conclusions. Recently, \citet{Lai2025} found that for an isotropic momentum distribution, the polarization degree, given by $\Pi= (p+1)/(p+7/3)$, remains invariant regardless of whether the electron energy distribution follows a power law, broken power law, double power law, kappa distribution, or log-parabolic distribution. In contrast, for anisotropic momentum distributions, this polarization formula is only valid in the high-frequency regime and breaks down at low frequencies. 

When self-absorption is taken into account, the optical depth is tightly correlated with the effective LOS integration depth. Specifically, different observational frequencies correspond to distinct effective emission depths, where SSA effects are non-negligible. Consequently, the radiation detected along the LOS is dominated by specific spatial regions with corresponding emission depths, rather than being uniformly contributed by the entire LOS path. This key property implies that multi-frequency polarization observations can be employed to probe the magnetic field structures in a layered manner—resolving the magnetic field distribution at different depths of the observed medium—thereby providing a promising approach for the three-dimensional diagnosis of MHD turbulence. Overall, this work explores the impacts of SSA on the polarization statistics of MHD turbulence in astrophysical environments. This framework is well-suited for application to multi-frequency data cubes obtained from state-of-the-art radio telescopes, such as the Low-Frequency Array (LOFAR) and the Square Kilometre Array (SKA). It enables quantitative investigations of MHD turbulence in astrophysical environments where SSA plays a significant role.

\section{Summary} 
\label{Sec_Sum}
We incorporated SSA into the radiative transfer of polarized synchrotron emission by means of 3D MHD turbulence simulations, thus extending polarization diagnostic analysis of MHD turbulence beyond the optically thin approximation. Our main findings are summarized as follows:

\begin{enumerate}
    \item The SSA turnover frequency $\nu_{\rm a}$, which corresponds to the spectral turnover point of synchrotron radiation with SSA, marks the transition of polarized emission from the optically thick to the optically thin regime. This frequency follows a power-law relation with magnetic field strength $B$ and emission integration depth $L$, given by $\nu_{\rm a} \propto (\kappa B^{(p+2)/2} L)^{2/(p+4)}$.

    \item SSA suppresses the amplitude of polarized intensity and homogenizes its spatial distribution, while also modifying the spectral energy distribution of polarized intensity. In the high-frequency optically thin regime, $P(\nu)\propto\nu^{-(p-1)/2}$ for all turbulence states; the distinct low-frequency power-law scalings arise from the combined effect of SSA and FR.

    \item SSA significantly weakens the anisotropy of polarized intensity, thereby suppressing the detectability of polarization anisotropy; FR alters the spatial distribution of polarized intensity, with moderate amplitude suppression and a slight reduction in polarization anisotropy.

    \item SSA preserves the canonical $k^{-8/3}$ power-law scaling of the polarized intensity statistics in the high-wavenumber inertial range, only reducing its overall amplitude. Although SSA alters the scaling relation between polarized intensity variance and wavelength and induces new characteristic behaviors (see Table \ref{Tab_Var}), this relation remains sensitive to the spectral index of magnetic turbulence, and thus can still be reliably used to diagnose the magnetic turbulence spectrum.
\end{enumerate}

This work demonstrates that while SSA profoundly shapes the observable polarization properties of turbulent astrophysical plasmas, it does not impede the diagnosis of MHD turbulence properties via polarization statistical techniques. These results extend the applicability of synchrotron polarization diagnostics to optically thick astrophysical environments (e.g., compact radio sources, relativistic jets of active galactic nuclei) and provide a crucial theoretical foundation for interpreting multi-frequency polarization observations from state-of-the-art radio telescopes such as LOFAR and the SKA.

\section*{Acknowledgements}
We would like to thank the anonymous referee for the valuable comments that improved our manuscript. We acknowledge the support from National SKA Program of China (2025SKA0140100), and the National Natural Science Foundation of China (grant Nos. 12473046, 12233002 and 12473042). 

\section*{Data Availability}
The data that support the findings of this study are available from the corresponding author upon reasonable request.

\bibliographystyle{mnras}
\bibliography{references} 

@ARTICLE{Burkhart2021,
       author = {{Burkhart}, Blakesley},
        title = "{Diagnosing Turbulence in the Neutral and Molecular Interstellar Medium of Galaxies}",
      journal = {\pasp},
         year = 2021,
        month = oct,
       volume = {133},
       number = {1028},
          eid = {102001},
        pages = {102001},
          doi = {10.1088/1538-3873/ac25cf},
archivePrefix = {arXiv},
       eprint = {2106.02239},
 primaryClass = {astro-ph.GA},
       adsurl = {https://ui.adsabs.harvard.edu/abs/2021PASP..133j2001B}
}

@BOOK{Rybicki1979,
       author = {{Rybicki}, George B. and {Lightman}, Alan P.},
        title = "{Radiative processes in astrophysics}",
         year = 1979,
       adsurl = {https://ui.adsabs.harvard.edu/abs/1979rpa..book.....R}
}

@ARTICLE{Waelkens2009,
       author = {{Waelkens}, A.~H. and {Schekochihin}, A.~A. and {En{\ss}lin}, T.~A.},
        title = "{Probing magnetic turbulence by synchrotron polarimetry: statistics and structure of magnetic fields from Stokes correlators}",
      journal = {\mnras},
         year = 2009,
        month = oct,
       volume = {398},
       number = {4},
        pages = {1970-1988},
          doi = {10.1111/j.1365-2966.2009.15231.x},
archivePrefix = {arXiv},
       eprint = {0903.3056},
 primaryClass = {astro-ph.GA},
       adsurl = {https://ui.adsabs.harvard.edu/abs/2009MNRAS.398.1970W}
}

@ARTICLE{Beck2015,
       author = {{Beck}, Rainer},
        title = "{Magnetic fields in spiral galaxies}",
      journal = {\aapr},
         year = 2015,
        month = dec,
       volume = {24},
          eid = {4},
        pages = {4},
          doi = {10.1007/s00159-015-0084-4},
archivePrefix = {arXiv},
       eprint = {1509.04522},
 primaryClass = {astro-ph.GA},
       adsurl = {https://ui.adsabs.harvard.edu/abs/2015A&ARv..24....4B}
}

@ARTICLE{LP12,
       author = {{Lazarian}, A. and {Pogosyan}, D.},
        title = "{Statistical Description of Synchrotron Intensity Fluctuations: Studies of Astrophysical Magnetic Turbulence}",
      journal = {\apj},
         year = 2012,
        month = mar,
       volume = {747},
       number = {1},
          eid = {5},
        pages = {5},
          doi = {10.1088/0004-637X/747/1/5},
archivePrefix = {arXiv},
       eprint = {1105.4617},
 primaryClass = {astro-ph.GA},
       adsurl = {https://ui.adsabs.harvard.edu/abs/2012ApJ...747....5L}
}

@ARTICLE{LP16,
       author = {{Lazarian}, A. and {Pogosyan}, D.},
        title = "{Spectrum and Anisotropy of Turbulence from Multi-frequency Measurement of Synchrotron Polarization}",
      journal = {\apj},
         year = 2016,
        month = feb,
       volume = {818},
       number = {2},
          eid = {178},
        pages = {178},
          doi = {10.3847/0004-637X/818/2/178},
archivePrefix = {arXiv},
       eprint = {1511.01537},
 primaryClass = {astro-ph.GA},
       adsurl = {https://ui.adsabs.harvard.edu/abs/2016ApJ...818..178L}
}

@ARTICLE{Zhang2019a,
       author = {{Zhang}, Jian-Fu and {Lazarian}, Alex and {Ho}, Ka Wai and {Yuen}, Ka Ho and {Yang}, Bo and {Hu}, Yue},
        title = "{Tracing magnetic field with synchrotron polarization gradients: parameter study}",
      journal = {\mnras},
         year = 2019,
        month = jul,
       volume = {486},
       number = {4},
        pages = {4813-4822},
          doi = {10.1093/mnras/stz1176},
archivePrefix = {arXiv},
       eprint = {1904.11939},
 primaryClass = {astro-ph.HE},
       adsurl = {https://ui.adsabs.harvard.edu/abs/2019MNRAS.486.4813Z}
}

@ARTICLE{Zhang2019b,
       author = {{Zhang}, Jian-Fu and {Liu}, Qian and {Lazarian}, Alex},
        title = "{Tracing Magnetic Fields By the Synergies of Synchrotron Emission Gradients}",
      journal = {\apj},
         year = 2019,
        month = nov,
       volume = {886},
       number = {1},
          eid = {63},
        pages = {63},
          doi = {10.3847/1538-4357/ab4b4a},
archivePrefix = {arXiv},
       eprint = {1910.02378},
 primaryClass = {astro-ph.HE},
       adsurl = {https://ui.adsabs.harvard.edu/abs/2019ApJ...886...63Z}
}

@ARTICLE{Lazarian2018,
       author = {{Lazarian}, A. and {Yuen}, Ka Ho and {Ho}, Ka Wai and {Chen}, Junda and {Lazarian}, Victor and {Lu}, Zekun and {Yang}, Bo and {Hu}, Yue},
        title = "{Distribution of Velocity Gradient Orientations: Mapping Magnetization with the Velocity Gradient Technique}",
      journal = {\apj},
         year = 2018,
        month = sep,
       volume = {865},
       number = {1},
          eid = {46},
        pages = {46},
          doi = {10.3847/1538-4357/aad7ff},
archivePrefix = {arXiv},
       eprint = {1802.02984},
 primaryClass = {astro-ph.GA},
       adsurl = {https://ui.adsabs.harvard.edu/abs/2018ApJ...865...46L}
}

@ARTICLE{Guo2024,
       author = {{Guo}, Ning-Ning and {Zhang}, Jian-Fu and {Xiao}, Hua-Ping and {Cho}, Jungyeon and {Yang}, Xue-Juan},
        title = "{Measurement of Interstellar Magnetization by Synchrotron Polarization Variance}",
      journal = {\apj},
         year = 2024,
        month = jun,
       volume = {967},
       number = {2},
          eid = {91},
        pages = {91},
          doi = {10.3847/1538-4357/ad452c},
archivePrefix = {arXiv},
       eprint = {2404.16310},
 primaryClass = {astro-ph.GA},
       adsurl = {https://ui.adsabs.harvard.edu/abs/2024ApJ...967...91G}
}

@ARTICLE{Zhang2025,
       author = {{Zhang}, Jian-Fu and {Liu}, Zi-Qi},
        title = "{Revealing the Spectral Properties of Magnetic Turbulence by Synchrotron Polarization Gradients}",
      journal = {\apj},
         year = 2025,
        month = jun,
       volume = {985},
       number = {2},
          eid = {235},
        pages = {235},
          doi = {10.3847/1538-4357/add5f2},
archivePrefix = {arXiv},
       eprint = {2505.04017},
 primaryClass = {astro-ph.HE},
       adsurl = {https://ui.adsabs.harvard.edu/abs/2025ApJ...985..235Z}
}

@ARTICLE{Wang2020,
       author = {{Wang}, Ru-Yue and {Zhang}, Jian-Fu and {Xiang}, Fu-Yuan},
        title = "{Studying the Anisotropy of Compressible Magnetohydrodynamic Turbulence by Synchrotron Polarization Intensity}",
      journal = {\apj},
         year = 2020,
        month = feb,
       volume = {890},
       number = {1},
          eid = {70},
        pages = {70},
          doi = {10.3847/1538-4357/ab6a1a},
archivePrefix = {arXiv},
       eprint = {2001.03276},
 primaryClass = {astro-ph.HE},
       adsurl = {https://ui.adsabs.harvard.edu/abs/2020ApJ...890...70W}
}

@ARTICLE{Wang2024,
       author = {{Wang}, Ru-Yue and {Zhang}, Jian-Fu and {Lu}, Fang and {Xiang}, Fu-Yuan},
        title = "{Exploring the intermittency of magnetohydrodynamic turbulence by synchrotron polarization radiation}",
      journal = {\aap},
         year = 2024,
        month = nov,
       volume = {691},
          eid = {A26},
        pages = {A26},
          doi = {10.1051/0004-6361/202450414},
archivePrefix = {arXiv},
       eprint = {2409.05739},
 primaryClass = {astro-ph.HE},
       adsurl = {https://ui.adsabs.harvard.edu/abs/2024A&A...691A..26W}
}

@ARTICLE{Burn1966,
       author = {{Burn}, B.~J.},
        title = "{On the depolarization of discrete radio sources by Faraday dispersion}",
      journal = {\mnras},
         year = 1966,
        month = jan,
       volume = {133},
        pages = {67},
          doi = {10.1093/mnras/133.1.67},
       adsurl = {https://ui.adsabs.harvard.edu/abs/1966MNRAS.133...67B}
}

@ARTICLE{Brentjens2005,
       author = {{Brentjens}, M.~A. and {de Bruyn}, A.~G.},
        title = "{Faraday rotation measure synthesis}",
      journal = {\aap},
         year = 2005,
        month = oct,
       volume = {441},
       number = {3},
        pages = {1217-1228},
          doi = {10.1051/0004-6361:20052990},
archivePrefix = {arXiv},
       eprint = {astro-ph/0507349},
 primaryClass = {astro-ph},
       adsurl = {https://ui.adsabs.harvard.edu/abs/2005A&A...441.1217B}
}

@ARTICLE{Zhang2018,
       author = {{Zhang}, Jian-Fu and {Lazarian}, Alex and {Xiang}, Fu-Yuan},
        title = "{Spectral Properties of Magnetohydrodynamic Turbulence Revealed by Polarization Synchrotron Emission with Faraday Rotation}",
      journal = {\apj},
         year = 2018,
        month = aug,
       volume = {863},
       number = {2},
          eid = {197},
        pages = {197},
          doi = {10.3847/1538-4357/aad182},
archivePrefix = {arXiv},
       eprint = {1807.01218},
 primaryClass = {astro-ph.HE},
       adsurl = {https://ui.adsabs.harvard.edu/abs/2018ApJ...863..197Z}
}

@ARTICLE{Zhang2016,
       author = {{Zhang}, Jian-Fu and {Lazarian}, Alex and {Lee}, Hyeseung and {Cho}, Jungyeon},
        title = "{Studying Magnetohydrodynamic Turbulence with Synchrotron Polarization Dispersion}",
      journal = {\apj},
         year = 2016,
        month = jul,
       volume = {825},
       number = {2},
          eid = {154},
        pages = {154},
          doi = {10.3847/0004-637X/825/2/154},
archivePrefix = {arXiv},
       eprint = {1605.01307},
 primaryClass = {astro-ph.HE},
       adsurl = {https://ui.adsabs.harvard.edu/abs/2016ApJ...825..154Z}
}

@ARTICLE{Xiao2025,
       author = {{Xiao}, Ya-Wen and {Zhang}, Jian-Fu and {Lazarian}, Alex and {Pogosyan}, Dmitri},
        title = "{Influence of Density Distribution on Synchrotron Polarization Dispersion in Magnetized Interstellar Medium}",
      journal = {\apj},
         year = 2025,
        month = sep,
       volume = {990},
       number = {2},
          eid = {169},
        pages = {169},
          doi = {10.3847/1538-4357/adf850},
archivePrefix = {arXiv},
       eprint = {2508.02938},
 primaryClass = {astro-ph.HE},
       adsurl = {https://ui.adsabs.harvard.edu/abs/2025ApJ...990..169X}
}

@ARTICLE{Xiao2026,
       author = {{Xiao}, Ya-Wen and {Zhang}, Jian-Fu and {Lazarian}, Alex},
        title = "{Revealing the Spectral Properties of the Galactic Interstellar Medium by Survey Observations}",
      journal = {\apj},
         year = 2026,
        month = apr,
       volume = {1001},
       number = {1},
          eid = {113},
        pages = {113},
          doi = {10.3847/1538-4357/ae5536},
archivePrefix = {arXiv},
       eprint = {2603.17464},
 primaryClass = {astro-ph.HE},
       adsurl = {https://ui.adsabs.harvard.edu/abs/2026ApJ..1001..113X}
}

@ARTICLE{Melrose1972,
       author = {{Melrose}, D.~B.},
        title = "{Variation in the Polarization Across Bends in the Spectra of Self-Absorbed Synchrotron Sources}",
      journal = {\pasa},
         year = 1972,
        month = oct,
       volume = {2},
       number = {3},
        pages = {140-142},
          doi = {10.1017/S132335800001328X},
       adsurl = {https://ui.adsabs.harvard.edu/abs/1972PASA....2..140M}
}

@ARTICLE{Jones1977,
       author = {{Jones}, T.~W. and {O'Dell}, S.~L.},
        title = "{Transfer of polarized radiation in self-absorbed synchrotron sources. I. Results for a homogeneous source.}",
      journal = {\apj},
         year = 1977,
        month = jun,
       volume = {214},
        pages = {522-539},
          doi = {10.1086/155278},
       adsurl = {https://ui.adsabs.harvard.edu/abs/1977ApJ...214..522J}
}

@ARTICLE{Ginzburg1965,
       author = {{Ginzburg}, V.~L. and {Syrovatskii}, S.~I.},
        title = "{Cosmic Magnetobremsstrahlung (synchrotron Radiation)}",
      journal = {\araa},
         year = 1965,
        month = jan,
       volume = {3},
        pages = {297},
          doi = {10.1146/annurev.aa.03.090165.001501},
       adsurl = {https://ui.adsabs.harvard.edu/abs/1965ARA&A...3..297G}
}

@ARTICLE{Cho2003,
       author = {{Cho}, Jungyeon and {Lazarian}, A.},
        title = "{Compressible magnetohydrodynamic turbulence: mode coupling, scaling relations, anisotropy, viscosity-damped regime and astrophysical implications}",
      journal = {\mnras},
         year = 2003,
        month = oct,
       volume = {345},
       number = {12},
        pages = {325-339},
          doi = {10.1046/j.1365-8711.2003.06941.x},
archivePrefix = {arXiv},
       eprint = {astro-ph/0301062},
 primaryClass = {astro-ph},
       adsurl = {https://ui.adsabs.harvard.edu/abs/2003MNRAS.345..325C}
}

@ARTICLE{Lazarian2025,
       author = {{Lazarian}, A. and {Ho}, Ka Wai and {Yuen}, Ka Ho and {Vishniac}, Ethan},
        title = "{Sub-Alfv{\'e}nic Turbulence: Magnetic-to-kinetic Energy Ratio, Modification of Weak Cascade, and Implications for Magnetic Field Strength Measurements}",
      journal = {\apj},
         year = 2025,
        month = jan,
       volume = {978},
       number = {1},
          eid = {88},
        pages = {88},
          doi = {10.3847/1538-4357/ad8d49},
archivePrefix = {arXiv},
       eprint = {2312.05399},
 primaryClass = {astro-ph.GA},
       adsurl = {https://ui.adsabs.harvard.edu/abs/2025ApJ...978...88L}
}

@ARTICLE{Ginzburg1969,
       author = {{Ginzburg}, V.~L. and {Syrovatskii}, S.~I.},
        title = "{Developments in the Theory of Synchrotron Radiation and its Reabsorption}",
      journal = {\araa},
         year = 1969,
        month = jan,
       volume = {7},
        pages = {375},
          doi = {10.1146/annurev.aa.07.090169.002111},
       adsurl = {https://ui.adsabs.harvard.edu/abs/1969ARA&A...7..375G}
}

@ARTICLE{Cowie2026,
       author = {{Cowie}, F.~J. and {Fender}, R.~P.},
        title = "{Towards improved synchrotron self-absorption energy estimates: accounting for inhomogeneous and non-spherical emitting regions}",
      journal = {\mnras},
         year = 2026,
        month = aug,
       volume = {550},
       number = {2},
          eid = {stag1113},
        pages = {stag1113},
          doi = {10.1093/mnras/stag1113},
archivePrefix = {arXiv},
       eprint = {2606.11307},
 primaryClass = {astro-ph.HE},
       adsurl = {https://ui.adsabs.harvard.edu/abs/2026MNRAS.550g1113C}
}

@ARTICLE{Carilli1991,
       author = {{Carilli}, C.~L. and {Perley}, R.~A. and {Dreher}, J.~W. and {Leahy}, J.~P.},
        title = "{Multifrequency Radio Observations of Cygnus A: Spectral Aging in Powerful Radio Galaxies}",
      journal = {\apj},
         year = 1991,
        month = dec,
       volume = {383},
        pages = {554},
          doi = {10.1086/170813},
       adsurl = {https://ui.adsabs.harvard.edu/abs/1991ApJ...383..554C}
}

@ARTICLE{Chevalier1998,
       author = {{Chevalier}, Roger A.},
        title = "{Synchrotron Self-Absorption in Radio Supernovae}",
      journal = {\apj},
         year = 1998,
        month = may,
       volume = {499},
       number = {2},
        pages = {810-819},
          doi = {10.1086/305676},
       adsurl = {https://ui.adsabs.harvard.edu/abs/1998ApJ...499..810C}
}

@ARTICLE{Corbel2002,
       author = {{Corbel}, S. and {Fender}, R.~P.},
        title = "{Near-Infrared Synchrotron Emission from the Compact Jet of GX 339-4}",
      journal = {\apjl},
         year = 2002,
        month = jul,
       volume = {573},
       number = {1},
        pages = {L35-L39},
          doi = {10.1086/341870},
archivePrefix = {arXiv},
       eprint = {astro-ph/0205402},
 primaryClass = {astro-ph},
       adsurl = {https://ui.adsabs.harvard.edu/abs/2002ApJ...573L..35C}
}

@ARTICLE{Sullivan2009,
       author = {{O'Sullivan}, S.~P. and {Gabuzda}, D.~C.},
        title = "{Magnetic field strength and spectral distribution of six parsec-scale active galactic nuclei jets}",
      journal = {\mnras},
         year = 2009,
        month = nov,
       volume = {400},
       number = {1},
        pages = {26-42},
          doi = {10.1111/j.1365-2966.2009.15428.x},
archivePrefix = {arXiv},
       eprint = {0907.5211},
 primaryClass = {astro-ph.CO},
       adsurl = {https://ui.adsabs.harvard.edu/abs/2009MNRAS.400...26O}
}

@ARTICLE{Lai2025,
       author = {{Lai}, Paul C.~W. and {Li}, Kaye J. and {Yap}, Y.~X. Jane and {Wu}, Kinwah and {Kong}, Albert K.~H.},
        title = "{Spectropolarimetry of synchrotron radiation from relativistic electrons with anisotropic pitch-angle and various energy distributions}",
      journal = {\mnras},
         year = 2025,
        month = sep,
       volume = {542},
       number = {2},
        pages = {902-916},
          doi = {10.1093/mnras/staf1295},
archivePrefix = {arXiv},
       eprint = {2509.04572},
 primaryClass = {astro-ph.HE},
       adsurl = {https://ui.adsabs.harvard.edu/abs/2025MNRAS.542..902L}
}

\bsp	
\label{lastpage}
\end{document}